\documentclass[aps,twocolumn,amsmath,amssymb,floatfix,superscriptaddress]{revtex4}
\usepackage{hyperref}
\usepackage{verbatim}
\usepackage{graphicx}
\usepackage{bm} 
\usepackage{dcolumn}  
\usepackage{color}

\begin{document}
\title{Pseudo-Casimir interactions across nematic films with disordered anchoring axis}

\author{Fahimeh Karimi Pour Haddadan}
\affiliation{Faculty of Physics, Kharazmi University, Karaj 31979-37551, Iran}

\author{Ali Naji}
\affiliation{School of Physics, Institute for Research in Fundamental Sciences (IPM), Tehran 19395-5531, Iran}

\author{Azin Kh. Seifi}
\affiliation{Faculty of Physics, Kharazmi University, Karaj 31979-37551, Iran}

\author{Rudolf Podgornik}
\affiliation{Department of Theoretical Physics, Jo\v{z}ef Stefan Institute, SI-1000 Ljubljana, Slovenia}
\affiliation{Department of Physics, Faculty of Mathematics and Physics, University of Ljubljana, SI-1000 Ljubljana, Slovenia}
\affiliation{Department of Physics, University of Massachusetts, Amherst, MA 01003, USA}

\begin{abstract}
We consider the effective pseudo-Casimir interaction forces mediated by a nematic liquid-crystalline film bounded by two planar surfaces, one of which imposes a random (disordered) distribution of the preferred anchoring axis in the so-called easy direction.  We consider both the case of a {\em quenched} as well as an {\em annealed} disorder for the easy direction on the disordered surface and analyze the resultant fluctuation-induced interaction between the surfaces. In the case of  quenched disorder, we show that the disorder effects appear additively in the total interaction and are  dominant at intermediate inter-surface separations. Disorder effects are shown to be unimportant at both very small and very large separations. In the case of annealed disorder its effects are non-additive in the total inter-surface interaction and can be rationalized in terms of a renormalized extrapolation length.
\end{abstract}

\maketitle

\section{Introduction}

Confined liquid crystals (LCs) have attracted much attention and are important from a fundamental as well as applied perspectives \cite{de-gennes}. Because of long-range (critical) correlations implied by the LC elastic Hamiltonian \cite{Kardar_rev}, the anchoring of the liquid crystal molecules adjacent to the confining substrates has a significant effect on the properties of the confined material. This effects can range from the changes wrought in the nature of the structural phase transitions of the confined LC material, as well as changes in the behavior of the correlation functions~\cite{hanke,radzihovsky}.
In a geometry where the director describing the medium deep in the LC phase is uniform and thus the elastic energy is zero, thermal fluctuations can show up through creating an effective inter-surface interaction \cite{LC,kardar}. These interactions have been dubbed {\em pseudo-Casimir} because of the similarities with electromagnetic (EM) Casimir-van der Waals interactions \cite{Casimir,Mostepanenko} also related to the long-range (critical) fluctuations of the EM fields \cite{LC}. While the phenomenology of the pseudo-Casimir interactions in confined LC slabs is quite rich \cite{rudi1,rudi2,rudi4,ziherl,karimi,karimi2} and can include interesting variation as in the case of nematic polymers \cite{rudijure}, they usually scale with the thermal energy, being consequently small and therefore not easily detectable.

However, in recent years there have been extensive theoretical investigations into model systems
that would more closely mimic real experimental conditions. In this respect models that consider natural inhomogeneities in samples' preparation affecting the shape of the boundary surfaces as well as the surface anchoring energy are particularly relevant and have been introduced in the context of EM Casimir interactions \cite{kardar} as well as in the context of confined LC slabs \cite{radzihovsky}. The boundary surface disorder induced by these inhomogeneities has been recently studied in great detail specifically within the context of Coulomb fluids \cite{Disorder1,Disorder2,partial,Moredisorder1,Moredisorder2,Moredisorder3,Moredisorder4,Moredisorder5}. They have been shown to cause a pronounced effect on the interactions between these disorder inducing surfaces mostly via the coupling between the thermal fluctuations and the surface induced disorder. This coupling leads, after an appropriate averaging depending on the nature of the disorder \cite{partial}, to additional free energy terms that depend on the separation between apposed surfaces. This disorder-induced interaction can be sometimes larger than the standard Casimir interaction and should thus be at least in principle easier to detect. While most of the works on disorder induced interactions have been limited to the Coulomb systems, there is no reason why the same type of phenomena should not be observable in other systems, characterized by critical correlations.

Here, we therefore investigate the effects of disorder in the homeotropic easy direction at the bounding surfaces of a nematic LC film \cite{west,shiy} on the interaction between them.  Describing the surface disorder fields with a simple Gaussian distribution for the homeotropic easy direction (with an assumed width or variance) allows us {\em grosso modo} to capture the orientational anchoring disorder present at heterogeneous surfaces. We consider both types of disorder: the quenched type as well as the annealed type and derive the corresponding  interaction free energies after the appropriate averaging over the disorder degrees of freedom. The coupling between the latter and the fluctuating fields is modelled within the
Rapini-Papoular surface interaction phenomenology \cite{Rapini}.
In the case of annealed disorder, we obtain an effective free energy for the fluctuating director field which takes a standard pseudo-Casimir form \cite{rudi1,rudi2,rudi4,ziherl,karimi,karimi2} but with a renormalized anchoring energy, while  in the case of quenched disorder, we find that the disorder in the homeotropic easy direction leads to a distinct additive term besides  the pseudo-Casimir contribution in the total interaction free energy. This is very similar to the general state of affairs in the context of Coulomb fluids under imposed external disorder in the monopolar charge distribution on bounding surfaces \cite{Moredisorder1}. In the present context, however, we show that the disorder effects are important only at an intermediate range of separation between the bounding surfaces and vanish for both very large and very small separations; this makes the easy direction disorder characteristically different from the charge disorder effects, which dominate only at large separations. The conclusion is then that the quenched disorder creates an  additive contribution, which is dependent on the disorder variance and can be thus as large as the disorder parameters allow it to be, whereas the annealed disorder leads to a non-additive modification of the fluctuation induced interaction between disordered bounding surfaces and is thus fundamentally limited in its magnitude even for large disorder variances.

The organization of the paper is as follows: In Section \ref{sec:model}, we present details of our model and the formalism employed to calculate the interaction free energy  due to the LC director fluctuations between disordered boundaries. In Section \ref{sec:quenched}, we analyze the results in the case of quenched disorder and then turn to the case of annealed disorder in Section \ref{sec:annealed}. We conclude our discussion in Section \ref{sec:concl}.

\section{Model and Formalism}
\label{sec:model}

We consider a nematic liquid-crystalline film in a hybrid cell geometry bounded by two flat, plane-parallel surfaces located at the positions $z=0$ and $z=d$ along the normal axis to the surfaces \cite{de-gennes}. The substrate at $z=0$ is assumed to impose a strong homeotropic anchoring, such  that the nematic phase can be characterized by a uniform mean director field ${\bf n}_0={\bf z}$. The preferred anchoring orientation of the director field, or the so-called easy direction ${\mathbf e}$, on the substrate at $z=d$ is assumed to be disordered as we shall specify later.  The strength of the anchoring on this substrate, denoted by
$W$, is assumed to be finite. The Frank's continuum  elastic  energy for the bulk phase is then given by
\begin{equation}
H_{{\mathrm b}}={1\over 2}\!\int d{\mathbf r}\, [K_1 (\nabla \cdot{\bf n})^2+K_2 ({\bf n}\cdot\nabla\times {\bf n})^2+K_3({\bf n}\times\nabla \times {\bf n})^2],
\label{bgcfhje}
\end{equation}
where ${\bf n}$ is the nematic director field, ${\mathbf r}=({\mathbf x}, z)$ with ${\mathbf x}=(x,y)$ being the lateral Cartesian coordinates, and $K_1$, $K_2$, and $K_3$ are the splay, twist, and bend elastic constants, respectively \cite{de-gennes}.

Here, we are interested in small fluctuations $\delta {\mathbf n}({\mathbf r})$ around the mean-field value of the director ${\mathbf n}_0({\mathbf r})=\hat {\mathbf z}$. Up to the second order, the fluctuation field can thus be written in terms of two fluctuation modes (corresponding to the massless modes that result from spontaneous breaking of two continuous rotational symmetries in nematics \cite{de-gennes}) as $\delta {\bf n}=(n_x, n_y, -1+\sqrt{1-n_x^2-n_y^2})\simeq (n_x, n_y, -n_x^2/2-n_y^2/2)$, and hence, ${\bf n}={\bf n}_0+\delta {\bf n}\simeq (n_x,n_y,1-n_x^2/2-n_y^2/2)$. Thus within the effective one-constant approximation, the bulk elastic energy Eq. (\ref{bgcfhje}) assumes the form
\begin{equation}
H_{{\mathrm b}}={K\over 2}\int d{\mathbf r}\,[(\nabla n_x)^2+(\nabla n_y)^2].
\end{equation}
The interaction of the nematic director with the substrate at $z=d$ is taken into account through the Rapini-Papoular surface energy model of a quadratic form \cite{Rapini}
\begin{equation}
H_{{\mathrm s}}=-{W\over 2}\int d^2 x\, ({\bf n}\cdot{\bf e})^2.
\end{equation}

The easy direction ${\bf e}({\mathbf x})$ is now assumed to have a  random distribution around the preferred or the mean easy direction ${\bf e}_0=\hat {\mathbf z}$. For the sake of consistency, we assume that the deviations from this preferred direction are small, i.e., the lateral components $e_x$ and $e_y$ are small while $e_z\simeq 1$. In spherical presentation  ${\bf e}=(\sin \theta \cos \phi,\sin \theta \sin \phi, \cos \theta)$, the latter approximation implies that the disordered easy direction exhibits small deviations in the polar angle $\theta$ from the constant homeotropic anchoring.
Thus,  at the leading order, the surface Hamiltonian, of this broken symmetry state (considering the fluctuations around the ${\bf n_0}=\hat {\bf z}$ ground state), reduces to
\begin{equation}
H_{{\mathrm s}}={W\over 2}\int d^2 x\, [n_x^2+n_y^2-2(n_x\, e_x +n_y\, e_y)],
\end{equation}
where  $e_i ({\bf x})= e_x({\bf x})$ and $e_y({\bf x})$ (for $i=x, y$) are treated as statistically independent fields with zero mean value $\langle e_i({\bf x})\rangle =0$ and equal two-point correlation function, i.e., $\langle e_i({\bf x}) e_j({\bf x}') \rangle = c({\bf x} - {\bf x}') \delta_{ij}$. The latter implies that disorder has a translationally invariant correlation. Specifically, we shall assume that the random distribution of these fields across the anchoring surface $z=d$ follows a Gaussian probability distribution function as
\begin{equation}
{\cal P}[e_i] = C \exp\Big\{-{1\over 2}\int d^{2}x \,d^{2}x' \, e_i({\bf x})c^{-1}({\bf x} - {\bf x}')e_i ({\bf x}')\Big\},
\end{equation}
where $C$ is a normalization constant.

For any given configuration of the easy direction fields, the partition function of the system can be calculated via
\begin{equation}
{\mathcal Z}[e_x, e_y]=\int\bigg(\prod_{i=x,y}{\mathcal D}n_i\bigg)\,  \exp(-\beta H[\{n_i\}; e_x, e_y]),
\label{eq:Z_1}
\end{equation}
where  $\beta=1/(k_BT)$ with $k_B$ and $T$ being the Boltzmann constant and the temperature, respectively,  and $H=H_{{\mathrm b}}+H_{{\mathrm s}}$ is the full Hamiltonian of the system. Clearly,  the partition function is a functional of the disordered easy direction fields and the free energy of the system is obtained after taking a proper ensemble average over all possible realizations of these fields. In order to do this, one should distinguish between different types of disorder, such as quenched or annealed disorder \cite{dotsenko1,dotsenko2}  (or even partially-annealed disorder \cite{partial} which will not be considered in this work).

In the {\em quenched disorder} model, the random distribution of the easy direction fields is frozen over the surface at $z=d$. Hence, one should calculate the average of  the sample free energy ${\mathcal F}[e_x, e_y]  = - k_BT  \ln {\mathcal Z}[e_x, e_y] $ over the disorder fields \cite{dotsenko1,dotsenko2} in order to obtain the thermodynamic free energy   of the system, i.e.,
\begin{equation}
  {\mathcal F} = - k_BT \langle \ln {\mathcal Z}[e_x, e_y] \rangle,
  \label{eq:quenched_av}
\end{equation}
where $\langle \cdots \rangle = \int \left(\prod_{i=x,y} {\mathcal D} e_i  {\mathcal P}[e_i]\right) (\cdots)$.
In the {\em annealed disorder} model, on the other hand, the disorder fields are assumed to be thermalized with the bulk degrees of freedom and should be thus treated on the same footing; hence, one should take the average of the partition function itself in order to obtain the free energy as
 \begin{equation}
  {\mathcal F} = - k_BT  \ln \langle{\mathcal Z}[e_x, e_y] \rangle.
 \end{equation}
In what follows, we shall first focus on the  quenched disorder model, which is the case of interest in this work, and then briefly consider also the case of annealed disorder.

\section{Quenched disorder}
\label{sec:quenched}

The disorder average in the quenched model,  Eq. (\ref{eq:quenched_av}), can be performed using the standard replica ``trick"  \cite{dotsenko1,dotsenko2}  by making use of the following {\em ansatz}:
\begin{equation}
\langle\ln {\mathcal Z}\rangle = \lim_{m\rightarrow 0}\frac{\langle {\mathcal Z}^m\rangle-1}{m} = \partial_{m}\langle{\mathcal Z}^m\rangle\big|_{m\rightarrow 0}.
\end{equation}
Note that the two  modes, $n_x$ and $n_y$,  are degenerate, and that the partition function (\ref{eq:Z_1}) can be factorized as ${\mathcal Z}[e_x, e_y] = \prod_{i=x,y} {\mathcal Z}[e_i]$. Thus,  since  the two easy direction fields are statistically uncorrelated and  have identical probability distribution functions, we can drop the subindex $i$ and in the end multiply the free energy by a factor of 2. After taking the Gaussian integrals over the disorder fields the   ``replicated" partition function is obtained  as
\begin{equation}
\langle {\mathcal Z}^m\rangle = \int \bigg(\prod_{\alpha=1}^m{\mathcal D}n_\alpha\bigg)\, \exp(-\beta H_{rep}[\{n_\alpha\}]),
\label{Z}
\end{equation}
where
\begin{widetext}
\begin{equation}
H_{rep}[\{n_\alpha\}] =   {\beta K\over 2} \sum_{\alpha} \int d{\mathbf r}\,(\nabla n_{\alpha})^2+{\beta W\over 2} \sum_{\alpha,\beta}\int d^2 x \, d^2 x'\, n_{\alpha}({\bf x})\Big(\delta_{\alpha\beta}\delta({\bf x}-{\bf x}')-{\beta W} c({\bf x}-{\bf x}')\Big)n_{\beta}({\bf x}'),
\end{equation}
\end{widetext}
which clearly couples different replicas of the system through the last term.

In order to proceed, we use the fact that the statistical properties of the system, as defined above, are translationally invariant in lateral directions and thus introduce the  Fourier transform for in-plane coordinates. In this way we find:
\begin{widetext}
\begin{equation}
H_{eff}[n({\bf q},z)]=\sum_{\alpha,\beta}\sum_{{\bf q}}\Big\{{KA\over 2}\int_{0}^{d}dz\, \big(|\partial_{z}n_{\alpha}({\bf q},z)|+q^2|n_{\alpha}({\bf q},z)|^2\big)\delta_{\alpha\beta}+{WA\over 2}n_{\alpha}\big({\bf q},d)(\delta_{\alpha\beta}-{\beta W} c({\bf q})\big)n_{\beta}^{*}({\bf q},d)\Big\},
\end{equation}
\end{widetext}
where $A$ is the surface area, ${\bf q}$ denotes the wave vector conjugate to lateral space ${\bf x}$, and $c({\bf q})$ is the Fourier transform of
$c({\bf x}-{\bf x}')$ given by $c({\bf q})=\int d^2 x\,  c({\bf x}-{\bf x}')\,e^{-i{\bf q}.({\bf x}-{\bf x}')}$. We shall further assume that the disordered easy direction fields have an isotropic
correlation function that implies  $c({\bf q})=c(q)$.

The replicated partition function Eq.~(\ref{Z}) can be reduced to
\begin{eqnarray}
&&\langle {\cal Z}^{m}\rangle=\prod_{{\bf q}}\int \left(\prod _{\alpha=1}^{m}{\cal D}n_{\alpha}({\bf q},d)\right)\times
\label{eq:Z_m}
\\
&&e^{-{\beta WA\over 2}\sum_{\alpha,\beta}n_{\alpha}({\bf q},d)\Big(\delta_{\alpha\beta}-{\beta W} c({\bf q})+{K\over W} q\coth(q d)\delta_{\alpha\beta}\Big)n_{\beta}^{*}({\bf q},d)},\nonumber
\end{eqnarray}
and can be evaluated straightforwardly by using the standard path-integral methods~\cite{rudi1,rudi2,rudi4,karimi,ziherl,kleinert} giving
\begin{equation}
{\mathcal Z}= e^{-\frac{1}{2} \ln \det G^{-1}},
\label{eq:Z_eff_2}
\end{equation}
where the elements of the matrix $G^{-1}$ for each mode ${\mathbf q}$ are given by
\begin{equation}
G^{-1}_{\alpha\beta}=\beta KA q\cosh(qd)\delta_{\alpha\beta}+\beta WA\Big(\delta_{\alpha\beta}-{\beta W} c(q)\Big)\sinh(qd).
\end{equation}
Taking logarithm of this expression ~\cite{petridis} then gives
\begin{eqnarray}
&&\ln G_{\alpha\beta}^{-1}=\ln\bigg(\beta KA q\cosh(qd)+\beta WA\sinh(qd)\bigg)\delta_{\alpha\beta}+\nonumber\\
&&\quad+{1\over m}\ln\Big(1-{m(\beta W)^2c(q)\sinh(qd)\over \beta K q\cosh(qd)+\beta W \sinh(qd)}\Big)I_{\alpha \beta},
\end{eqnarray}
where $I_{\alpha \beta}$ is a matrix with all elements equal to one.

Finally, the total free energy of the system is obtained as
\begin{eqnarray}
\beta {\cal F}&=&\sum_{\bf q}\Big[\ln\bigg(\beta KA q\cosh(qd)+\beta WA\sinh(qd)\bigg)\nonumber\\
&-&{(\beta W)^2 c(q)\sinh(qd)\over \beta K q\cosh(qd)+\beta W\sinh(qd)}\Big],
\end{eqnarray}
where the contribution of  both director modes has been taken into account.

It is evident that the  free energy of the system consists of two additive terms $\beta {\cal F}=\beta
{\cal F}_0+\beta {\cal F}_{{\mathrm{dis}}}$, where ${\cal F}_0$ is the standard pseudo-Casimir
interaction due to nematic fluctuations in a film (bounded by a surface with infinite anchoring on $z=0$ and and another surface with finite anchoring at $z=d$)~\cite{rudi1}
and the second term, ${\cal F}_{{\mathrm{dis}}}$, is due to the quenched statistics of the easy direction (on the $z=d$ surface).
The explicit forms of the pseudo-Casimir term is given by
\begin{equation}
{{\cal F}_0\over A}={k_B T\over 2\pi}\int_{0}^{\infty}{q\ln \Big(
1+{q-\ell^{-1}\over q+\ell^{-1}}\exp(-2qd)}\Big)dq,
\end{equation}
while the disorder contribution is obtained as
\begin{equation}
{{\cal F}_{{\mathrm{dis}}}\over A}=-{K\over 2\pi \ell^2}\int_{0}^{\infty}{c(q)q
~dq\over q\coth(qd)+\ell^{-1}}, \label{fcorr}
\end{equation}
where $\ell=K/W$ is the {\em extrapolation length } and we have used the continuum representation of the Fourier mode summation by using $\sum_{\bf q}=A\int d^2q/(2\pi)^2$. Also the irrelevant terms due to the free energies of the bulk and surfaces are omitted from the free energy.

We thus find that the effects of quenched disorder in the easy direction  appear in an {\em additive} form in the free energy. This closely resembles the behavior seen in Coulomb fluids bounded by disordered charge distributions \cite{Disorder1,Disorder2} or indeed also {\em in vacuo} between disordered charge distributions \cite{Moredisorder1,Moredisorder2,Moredisorder3,Moredisorder4,Moredisorder5}, where the quenched disorder effects also appear in an additive form. This property however does not hold in general; for instance, for a nematic film with quenched disordered anchoring energy, we find a more complicated non-additive behavior which will be discussed elsewhere \cite{paper1}. Also we note that the disorder contribution to the free energy Eq. (\ref{fcorr}) is intrinsically not limited in magnitude and can vary depending on the disorder parameters. This is fundamentally different from the annealed case that exhibits an intrinsic upper bound (see below).

\subsection{Effective interactions}

The effective interactions between bounding surfaces mediated by nematic fluctuations are thus modified by an additive contribution when the easy direction exhibits a quenched random distribution on one of the anchoring surfaces. The corresponding inter-surface force can be obtained by differentiating the free energy with respect to the inter-surface separation, $d$, as
\begin{equation}
f = -\partial {\cal F}/\partial d = f_0+f_{\mathrm{dis}}.
\label{eq:f_f0_fdis}
\end{equation}
The pseudo-Casimir force $f_0$  reads
\begin{equation}
{f_{0}\over A}={k_B T\over
\pi}\int_{0}^{\infty}{q^2dq\over 1+{q+\ell^{-1}\over q-\ell^{-1}}e^{2qd}},
\label{force_0}
\end{equation}
while the force due to the quenched disorder follows as
\begin{eqnarray}
{f_{{\mathrm{dis}}}\over A}&&={K\over 2\pi \ell^{2}}\int_{0}^{\infty}{c(q)q^3
dq\over \Big[q\cosh(qd)+\ell^{-1}\sinh(qd)\Big]^2} \\
&&={2K\over
\pi \ell^{2}}\int_{0}^{\infty}{c(q)q^3 e^{2 q d}dq\over (1+{q+\ell^{-1}\over q-\ell^{-1}}e^{2qd})^{2} (q-\ell^{-1})^2}.
\label{eq:f_dis}
\end{eqnarray}

In order to determine the asymptotic behavior of the interaction force, we shall first consider the special case of an uncorrelated disorder
with  $c({\bf x}-{\bf
x}')=c_{0}\delta ({\bf x}-{\bf x}')$, or in Fourier representation $c(q)=c_0$, where
$c_0$ is the variance of the easy direction distribution, which  has  a dimension of $({\textrm{length}})^2$.

In the limit of small inter-surface separations, $d\ll \ell$ (or relatively large extrapolation lengths
$\ell\rightarrow\infty$, corresponding to the ``weak disorder" regime),  the integrations in the force expressions can be performed explicitly.
Hence, for the pseudo-Casimir force, Eq.~(\ref{force_0}), we find  the standard universal form
\begin{equation}
{{f}_{0}(d\ll \ell)\over A}=k_B T\frac{3\zeta (3)}{16\pi d^3},
\end{equation}
where $\zeta (3)=1.2020569\cdots$.
The force is repulsive
and diverges as $1/d^3$ as $d \rightarrow 0$. The force generated by the disorder, Eq.~(\ref{eq:f_dis}), is obtained in this case as
\begin{equation}
{f_{{\mathrm{dis}}}(d\ll \ell)\over A}={Kc_{0}\ln 2\over 2\pi\ell^{2}d^{2}},
\end{equation}
which is clearly non-universal and depends on the disorder variance and the elastic constant $K$ (the latter may be expressed in terms of the  molecular
length $a$ as $K\sim k_{B}T_{NI}/a$ where $T_{NI}$ is the temperature of isotropic-nematic phase transition~\cite{de-gennes}). The disorder force is thus repulsive as well but diverges for small separations as $1/d^2$. The disorder effects are therefore expected to
become important for sufficiently large separations
\begin{equation}
d> d_l \equiv \frac{3\zeta(3)k_BT\ell^2}{8Kc_0\ln 2}.
\end{equation}

In the limit of large inter-surface separations, $d\gg \ell$ (or $\ell\rightarrow 0$ corresponding to the ``strong disorder" regime),  the pseudo-Casimir contribution falls off in a universal fashion exhibiting an attractive force behavior as
\begin{equation}
{{f}_{0}(d\gg \ell)\over A}=- k_B T\frac{\zeta (3)}{4\pi d^3}.
\end{equation}
The disorder contribution in this regime remains repulsive and decays as
\begin{equation}
{f_{{\mathrm{dis}}}(d\gg \ell)\over A}={3\zeta(3)K c_{0}\over 4\pi d^4}.
\end{equation}

It thus turns out that the disorder would be dominant for separations
\begin{equation}
d < d_u \equiv  \frac{3Kc_{0}}{k_BT}.
\end{equation}

\begin{center}
\begin{figure}[t!]
\includegraphics[width=8.cm,angle=0]{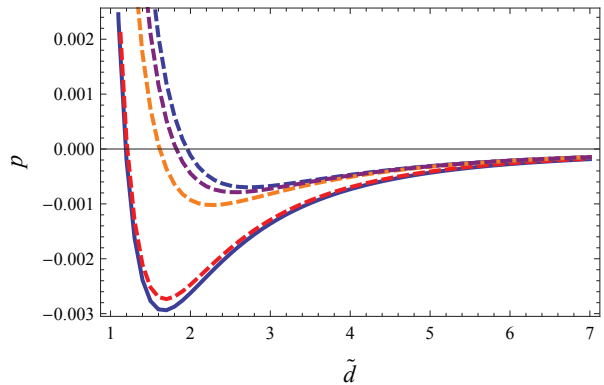}
\caption{The rescaled total force per unit area $p=\ell^3\beta f/A$ is shown as a function of the rescaled inter-surface separation ${\tilde d}=d/\ell$
for the case of a {\em quenched}
easy direction disorder with variance  $c_0=(k_B T/2K)\ell$ and correlation length $\xi/\ell=0, 0.5, 1, 10$ from top to bottom. The solid curve shows the rescaled disorder-free force, Eq.~(\ref{force_0}).}
\label{fig3}
\end{figure}
\end{center}

From the foregoing discussion, we conclude that the effects of quenched, uncorrelated easy direction  disorder are expected to be relevant in the {\em intermediate} regime of separations $d_l < d < d_u$.
This behavior is shown in Fig. \ref{fig3} (top curve) where we have used  $c_0 = (k_B T/2K)\ell\sim a\ell/2$. According to the above estimate, the effect of disorder dominates for $0.4 < d/\ell < 1.5$ (note that the extrapolation length can vary in a wide range of values, e.g., $\ell\sim a-10^3a$, where $a\sim 1$~nm is the molecular size).
As seen in the Figure, there is a crossover zero-force separation above (below) which the total force is attractive (repulsive); this is in fact the distance where the two surfaces form a stable equilibrium {\em bound state}.

It is also interesting to note that  the presence of  easy direction disorder {\em strengthens} the total interaction force at small separations, shifts the crossover distance to larger values (and thus tends to de-stabilize the bound state), and {\em suppresses} the attractive regime at intermediate to large inter-surface distances.

Note that, in general, the disorder in the easy direction may display a finite correlation length $\xi$~\cite{terentjev}. As a simple model, we can adopt a short-ranged correlation function
$c({\bf x}-{\bf x}')=c_{0}K_{0}({\bf x}/\xi)/(2\pi\xi^{2})$, where $K_0$ is the zeroth-order modified
Bessel function of the second kind.  In Fourier representation, the disorder correlation function takes a Lorentzian form
\begin{equation}
c(q)={c_0\over \xi^{2} q^{2}+1}.
\label{eq:c_q}
\end{equation}
The effects of disorder correlation can be examined by inserting this form of the correlation function in Eqs. (\ref{force_0})-(\ref{eq:f_dis}). As seen in Fig.~\ref{fig3},
the strength of the (repulsive) disorder-induced force diminishes as the disorder correlation length is increased. As a result, the bound-state between the two surfaces occurs at smaller  separations.

\begin{center}
\begin{figure}[t!]
\includegraphics[width=8.cm,angle=0]{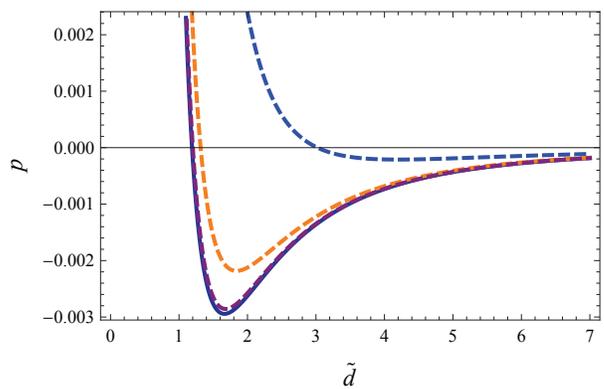}
\caption{The rescaled total force per unit area $p=\ell^3\beta f/A$ is shown as a function of the rescaled inter-surface separation ${\tilde d}=d/\ell$
for the case of a {\em quenched}
uncorrelated
easy direction disorder $\xi=0$ with different rescaled disorder variances  $\beta K c_0/\ell=0, 0.01, 0.1, 1$ from bottom to top. The solid curve shows the rescaled disorder-free force, Eq.~(\ref{force_0}).}
\label{fig3'}
\end{figure}
\end{center}

A similar trend is found when the variance of the easy direction disorder is varied. As seen in Fig.~\ref{fig3'}, upon decreasing the variance $c_0$ the effective interaction force decreases and approaches the disorder-free limit.

\section{Annealed disorder}
\label{sec:annealed}

As noted before, in the case of annealed disorder, the free energy is calculated from the relation ${\mathcal F} = - k_BT  \ln \langle{\mathcal Z}[e_x, e_y] \rangle$. The disorder-averaged
partition function $\langle{\mathcal Z}[e_x, e_y] \rangle$ follows simply by setting $m=1$ in Eq. (\ref{eq:Z_m}). The total free energy of the system in the annealed case is thus obtained as
\begin{equation}
\beta {\cal F}=\sum_{\bf q} \ln\bigg(\beta KA q\cosh(qd)+\beta W_{eff}(q)A\sinh(qd)\bigg),
\end{equation}
where the disorder leads to a renormalization of the anchoring energy to an effective form given by
$$W_{eff}(q)  = W(1-{\beta W}c(q)).$$This translates furthermore to an effective $q$-dependent extrapolation length given by
\begin{equation}
{1\over \ell_{eff}(q)}={1\over \ell}\Big(1-{\beta K\over \ell}c(q)\Big).
\label{eq:leff}
\end{equation}
The total force between the surfaces is then obtained straightforwardly as
\begin{equation}
{f\over A}={k_BT\over \pi} \int_{0}^{\infty} {{q^2 dq }\over
{1+{{q+\ell^{-1}_{eff}(q)}\over{q-\ell^{-1}_{eff}(q)}} e^{2 q d}}}.
\label{cfhjw}
\end{equation}
Obviously the result for annealed disorder significantly differs from that obtained in the case of quenched disorder. In the latter case, the disorder leads to an additive contribution to the total free energy (Eq. (\ref{eq:f_f0_fdis})), while in the former case, we find a non-additive form, where the disorder effects are inseparable from the pseudo-Casimir effects.
\begin{center}
\begin{figure}
\includegraphics[width=8.cm,angle=0]{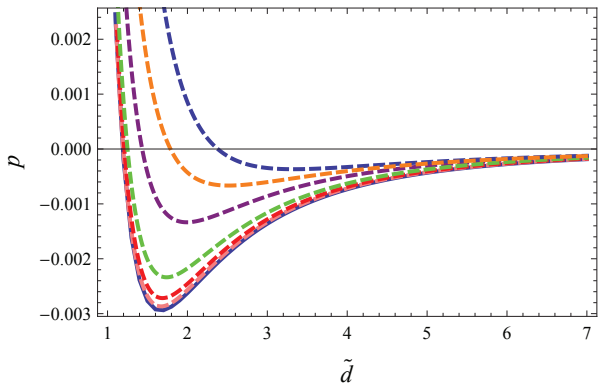}
\caption{The rescaled total force per unit area $p=\ell^3\beta f/A$ is shown as a function of the rescaled inter-surface separation ${\tilde d}=d/\ell$
for the case of an {\em annealed}
easy direction disorder with variance  $c_0=(k_B T/2K)\ell$ and correlation length $\xi/\ell=0, 1, 2, 5, 10, 20$ from top to bottom.
The solid curve shows the rescaled disorder-free force, Eq.~(\ref{force_0}).}
\label{fig4}
\end{figure}
\end{center}

Also, as is obvious from the above expression, the force is bounded by two limits characterized by $\ell_{eff}(q) \rightarrow \infty$ and $\ell_{eff}(q) \rightarrow 0$. In both cases the force per unit area is finite and we then in general have
\begin{equation}
\frac{3 \zeta(3)}{16{\pi d^3}} > {\beta f\over A} > -\frac{\zeta(3)}{4 {\pi d^3}}.
\label{eq:limits_a}
\end{equation}
In the context of the EM thermal Casimir effect the lower bound would correspond to the case of an ideally polarizable, i.e. metallic, surface.
Casimir interactions between metals are indeed an upper bound for the strength (magnitude) of the Casimir interactions proper. In general Eq. (\ref{cfhjw}) is of exactly the same form as obtained in the EM case for surfaces carrying a correlated dipolar layer, see Ref. \cite{rudi5}.
\begin{center}
\begin{figure}
\includegraphics[width=8.cm,angle=0]{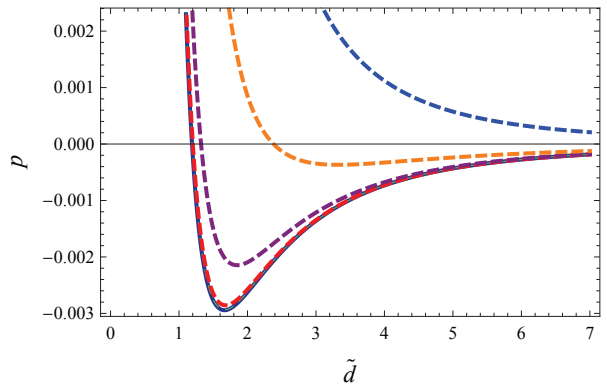}
\caption{The rescaled total force per unit area $p=\ell^3\beta f/A$ is shown as a function of the rescaled inter-surface separation ${\tilde d}=d/\ell$
for the case of an {\em annealed} uncorrelated easy direction disorder $\xi=0$ with different rescaled disorder variances  $\beta K c_0/\ell =0, 0.001, 0.01, 0.1, 0.5, 1$ from bottom to top.
The solid curve shows the rescaled disorder-free force, Eq.~(\ref{force_0}).}
\label{fig4'}
\end{figure}
\end{center}

\begin{center}
\begin{figure}
\includegraphics[width=8.cm,angle=0]{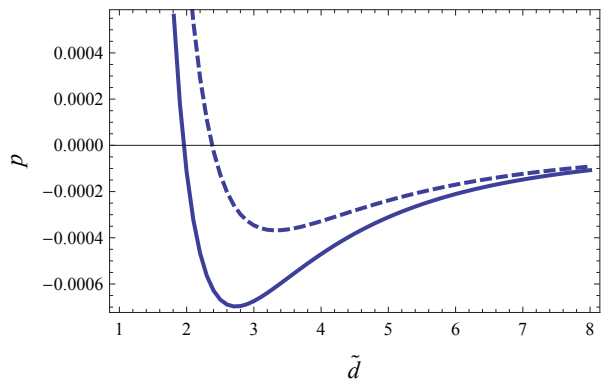}
\caption{The rescaled total force per unit area $p=\ell^3\beta f/A$ as a function of the rescaled inter-surface separation ${\tilde d}=d/\ell$ is shown for the two cases of uncorrelated annealed (top) and quenched  (bottom) disorder in the easy direction
with variance $c_0=(k_B T/2K)\ell$.} 
\label{fig5}
\end{figure}
\end{center}

The strengths of the disorder contribution again decreases upon increasing the disorder correlation, Fig. \ref{fig4}, and decreasing the variance, Fig.~\ref{fig4'}. It is however interesting to note that, for a given set of parameters,
the total interaction force in the annealed case is found to be more repulsive (less attractive) than that in the quenched case (see Fig. \ref{fig5}).

\section{Conclusion}
\label{sec:concl}

In this work, we have investigated the effective fluctuation-induced force between two planar boundaries that confine a nematic liquid crystalline film in the situation where one of the bounding surfaces imposes a random distribution for the preferred anchoring axis (easy direction) of the nematic director field. We have treated both cases of quenched (with a frozen disordered distribution for the easy axis) and annealed disorder (with a random but thermalized distribution of the easy direction). It is shown that in the quenched case the total interaction free energy is decomposed into two separate (additive) contributions, namely, a standard disorder-free pseudo-Casimir interaction and a disorder-induced contribution which is proportional to the variance of the disorder distribution. This latter quantity can be tuned to enhance or suppress the disorder effects independently of the pseudo-Casimir effect. On the contrary, in the annealed case the disorder effects can not be dissociated from the pseudo-Casimir interaction so that the whole effect can be expressed by an effective pseudo-Casimir interaction free energy but with a renormalized extrapolation length. The free energy in this case is bounded by two standard universal limiting laws [see Eq. (\ref{eq:limits_a})] and, therefore, the disorder effect (for weak or strong disorder variance) is also bracketed by these limiting laws.

In both cases of annealed and quenched disorder, we have shown that the easy direction disorder leads to a more repulsive interaction force  at small separations as compared with the (disorder-free) pseudo-Casimir force.  Also it leads to a shift to larger values of the crossover position where the force vanishes, that is, the stable bound-state between the surfaces becomes less stable.

We should also note that according to Eq. (\ref{eq:leff}) the effective extrapolation length can in principle become negative unless $c(q) \leq c_{cr} $, where $c_{cr} = \ell/\beta K$ is the critical strength of the disorder correlations. This implies that there is a regime in the parameter space where disorder completely destabilizes the system and is thus outside the confines of the methodology of the present theoretical framework. For the simple case of an uncorrelated disorder $c(q) = c_0$, or in general for $c(q)$ of the type shown in Eq. (\ref{eq:c_q}), the destabilizing effect of disorder is avoided for all modes provided that  $c_0\leq c_{cr}$. If we then use $K\sim k_BT/a$, we end up with $c_0\lesssim a\ell$. As $c_0$ measures the variance of the disorder in the easy direction this condition gives an estimate of the maximum polar angle fluctuations of the easy direction that would still not destabilize the system and is given by $\langle \theta^2\rangle = c_0/a^2 \lesssim \ell/a$. This is a reasonable assumption since $\ell/a$ can take a wide range of non-zero values for a finite anchoring. We therefore expect that the requirements for the positivity of the effective extrapolation length, giving rise to a stable solution in the presence of disorder, holds in most realistic situations.

While characteristic properties of disorder induced interactions in confined LCs are similar to the general behavior found in systems with  monopolar charge disorder in the context of EM Casimir effect or Coulomb fluids \cite{Disorder1,Disorder2,partial,Moredisorder1,Moredisorder2,Moredisorder3,Moredisorder4,Moredisorder5}, our results nevertheless  show a fundamental difference between this latter case and the confined LCs. The easy direction disorder turns out to become important in general only in an intermediate range of inter-surface separations-- or, in other words, for intermediate strength of the anchoring energy-- and becomes negligible (relative to the pseudo-Casimir contribution) both for very small and very large separations. This is obviously contrary to the behavior of a disordered Coulomb system, where the effects of charge disorder  become gradually more important as the disorder builds up and eventually dominate at large separations. Our results thus clearly point to differences between the disorder effects in these two types of systems that are both characterized by critical correlations.  The disorder effects are therefore not portable outside the exact nature of the system even if thermal and disorder correlations have the same qualitative behavior.

\begin{acknowledgments}
R.P. acknowledges ARRS grants J1-4297 and P1-0055.
\end{acknowledgments}

\end{document}